\documentclass[conference,10pt]{IEEEtran}
\IEEEoverridecommandlockouts
\usepackage{cite}
\usepackage{comment}
\usepackage{amsmath,amssymb,amsfonts}
\usepackage{algorithmic}
\usepackage{graphicx}
\usepackage{textcomp}
\usepackage{xcolor}
\usepackage{hyperref}
\usepackage[flushleft]{threeparttable}
\def\BibTeX{{\rm B\kern-.05em{\sc i\kern-.025em b}\kern-.08em
    T\kern-.1667em\lower.7ex\hbox{E}\kern-.125emX}}

\begin{document}
\bstctlcite{IEEEexample:BSTcontrol}

\title{Hardware-Software Co-design for Distributed Quantum Computing}

\author{
\IEEEauthorblockN{
Ji Liu\IEEEauthorrefmark{1}\IEEEauthorrefmark{4},
Allen Zang\IEEEauthorrefmark{2}\IEEEauthorrefmark{4},
Martin Suchara\IEEEauthorrefmark{3},
Tian Zhong\IEEEauthorrefmark{2}, and
Paul D Hovland\IEEEauthorrefmark{1}
}

\IEEEauthorblockA{
\IEEEauthorrefmark{1}Argonne National Laboratory~~\IEEEauthorrefmark{2}The University of Chicago~~\IEEEauthorrefmark{3}Microsoft Azure Quantum~~\IEEEauthorrefmark{4}Equal contribution\\
\{ji.liu,~hovland\}@anl.gov, \{yzang,~tzh\}@uchicago.edu, msuchara@microsoft.com
}
}

\maketitle

\begin{abstract}
    Distributed quantum computing (DQC) offers a pathway for scaling up quantum computing architectures beyond the confines of a single chip. Entanglement is a crucial resource for implementing non-local operations in DQC, and it is required to allow  teleportation of quantum states and gates. Remote entanglement generation in practical systems is probabilistic, has longer duration than that of local operations, and is nondeterministic. Therefore, optimizing the performance of probabilistic remote entanglement generation is critically important for the performance of DQC architectures. In this paper we propose and study a new DQC architecture that combines (1) buffering of successfully generated entanglement, (2) asynchronously attempted entanglement generation, and (3) adaptive scheduling of remote gates based on the entanglement generation pattern. We show that our hardware-software co-design improves both the runtime and the output fidelity under a realistic model of DQC.
\end{abstract}

\section{Introduction}\label{sec:intro}
Quantum computing is able to efficiently solve important classes of computational problems that are hard to solve using classical computers~\cite{shor1999polynomial, lanyon2010towards, farhi2014qaoa}. To make the computational power of quantum computers beneficial for real-world applications, quantum computing platforms need to be scaled beyond the size of state-of-the-art systems with limited number of qubits that have been demonstrated so far. However, placement of a large number of qubits within one monolithic quantum processing unit is expected to result in system performance degradation due to cross-talk between nearby physical qubits, limited capability of storing the physical qubits on a single chip with a small footprint, and increased complexity of controlling the individual qubits. Scaling up monolithic QPUs will be therefore constrained by various technical obstacles that are difficult to overcome.

In contrast to monolithic QPUs, distributed quantum computing (DQC)~\cite{cuomo2020towards,caleffi2024distributed,barral2024review} is a paradigm which aims to interconnect multiple monolithic QPUs together with the goal to allow evaluation of quantum circuits whose sizes exceed the number of qubits in a single QPU. DQC is viewed as an important milestone in scaling quantum computers, as described in the roadmaps of many quantum hardware manufacturers~\cite{carrera2024IBM_DQC, afzal2024distributed, IonQRemoteIonIonEntanglement}. DQC allows finding the sweet spot for the size of the individual QPUs, and therefore mitigates the aforementioned monolithic QPU scaling difficulties. At the same time, DQC introduces new challenges. Entanglement between the QPUs is the most important resource in DQC, which is consumed to implement remote multi-qubit gates. The generation of remote entanglement involves photon transmission, photonic Bell state measurement for heralding of success, and feedforward of measurement results. These processes result in a probabilistic success of remote entanglement generation and a longer cycle time for each attempt compared to local quantum operations. These entanglement generation features represent a significant bottleneck in the DQC performance. Implementation of remote gates requires prior successful entanglement generation, leading to potential gate delays that may further cascade through the system due to gate dependencies. When initialized qubits sit idle, this introduces additional decoherence leading to circuit output fidelity degradation as well as slower evaluation of the circuit. 

In this paper, we study a new DQC architecture which incorporates hardware-software co-design. The three key principles of our architecture are:
\begin{enumerate}
    \item Use of three different types of hardware qubits: communication qubits which are used for remote entanglement generation, buffer qubits which store the successfully generated entangled states, and data qubits which are used for quantum circuit evaluations.
    \item Asynchronous attempts to generate remote entanglement. This smoothens the temporal pattern of successfully generated entanglement, leading to better time resource management.
    \item Division of the quantum circuit into segments and identification of equivalent variants of circuit segments obtained by commuting remote gates to allow more efficient adaptive scheduling of remote gates by using information about the success of entanglement generation.
\end{enumerate}

This paper is organized as follows: In Sec.~\ref{sec:background} we review the necessary background and survey related work. We describe our architectural framework in Sec.~\ref{sec:framework}, and the evaluation methodology in Sec.~\ref{sec:methodology}. Our evaluation results are presented in Sec.~\ref{sec:results}. Finally, we conclude in Sec.~\ref{sec:conclusion}.

\section{Background}\label{sec:background}
In this section, we review basics of quantum information, the mechanism of heralded remote entanglement generation, and the implementation of basic non-local operations in DQC. Then we review relevant literature.

\subsection{Quantum information basics}
Quantum information is encoded in continuous complex probability amplitudes. Noiseless (pure) quantum states are linear combinations of eigenstates of the physical systems, i.e. quantum superpositions. Quantum states can be written as $|\psi\rangle = \alpha|0\rangle + \beta|1\rangle$ with $\alpha,\beta\in\mathbb{C}$, where $|0\rangle$ and $|1\rangle$ are orthonormal bases of a two-dimensional Hilbert space, and normalization requires $|\alpha|^2+|\beta|^2=1$. In quantum computing, an entire quantum circuit is a unitary operator $U$, s.t. $U^\dagger U=UU^\dagger=I$, where the dagger in the superscript denotes composition of complex conjugation and transpose, and $I$ is the identity operator. 

Entanglement is another key feature of quantum mechanics, and also one of the most important resources in DQC. A canonical example of entangled states are the 4 orthonormal \textit{Bell states (EPR pairs)} of two qubits: $|\Phi^{\pm}\rangle = \frac{1}{\sqrt{2}}(|00\rangle\pm|11\rangle)$, $|\Psi^{\pm}\rangle = \frac{1}{\sqrt{2}}(|01\rangle\pm|10\rangle)$. The Bell states can be transformed into each other via applications of single-qubit operations known as \textit{Pauli operators} which are defined as follows: $X|0\rangle=|1\rangle$, $X|1\rangle=|0\rangle$ (also denoted as $\oplus$), $Z|0\rangle=|0\rangle$, $Z|1\rangle=-|1\rangle$, and $Y=iXZ$.

Environment perturbations will cause decoherence by imposing errors on quantum states. Decoherence generally transforms quantum states to mixed states, i.e. a statistical mixture of pure quantum states. A common error model is the Pauli channel that applies each of the three Pauli operators and the identity operator with certain probabilities. The \textit{fidelity} of the mixed output quantum state $\rho$ with respect to the ideal pure output state $|\psi\rangle$: $F=\langle\psi|\rho|\psi\rangle$, is a key performance metric.

\begin{figure}[t]
    \centering
    \includegraphics[width=0.85\linewidth]{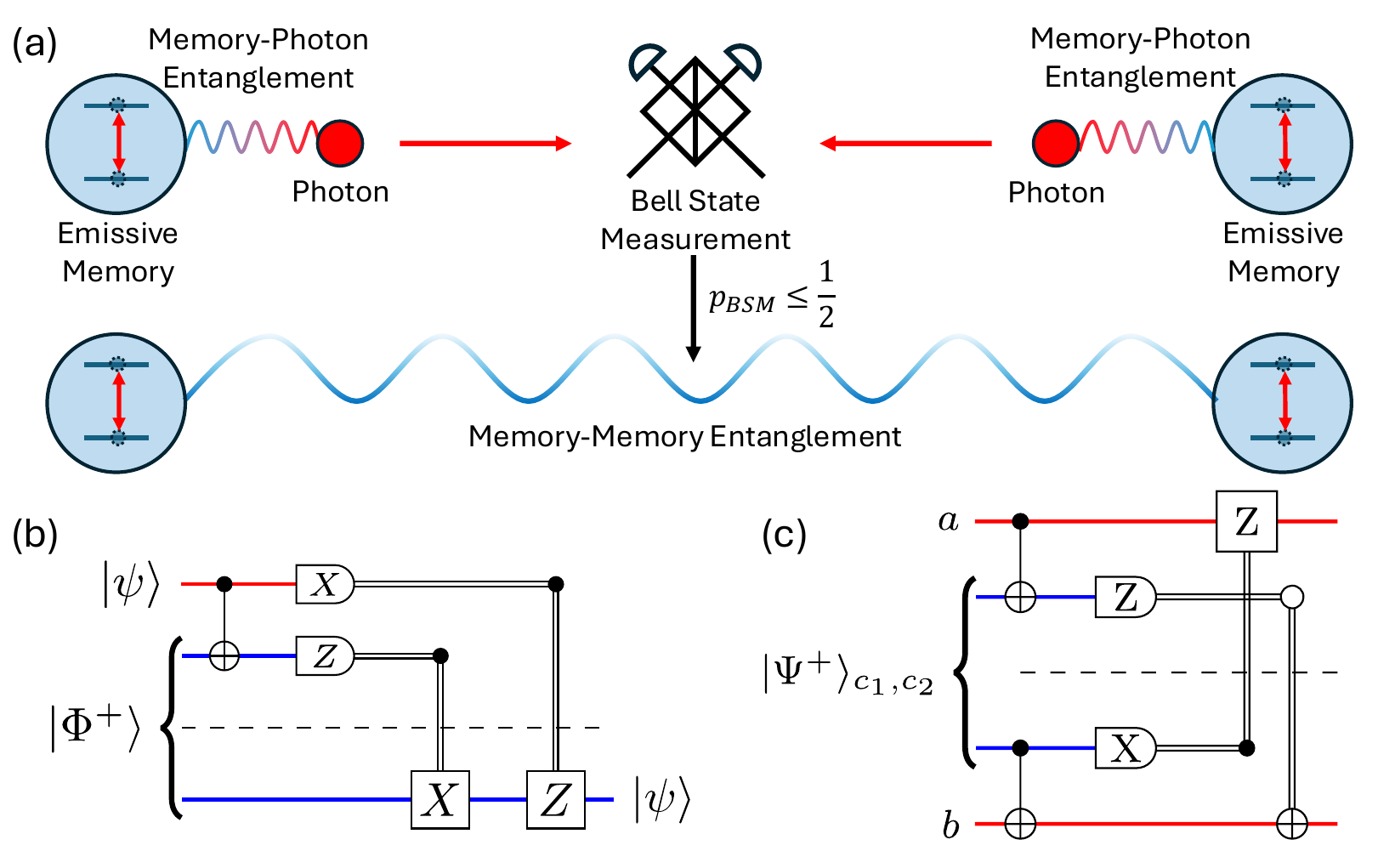}
    \caption{Preliminaries of DQC. (a) Heralded remote entanglement generation. (b) Quantum state teleportation circuit. (c) CNOT teleportation circuit. The dashed lines denote the partition of two local parties. The blue lines represent the Bell pair resource and the red lines represent data qubits.}
    \label{fig:preliminaries}
\end{figure}

\subsection{Heralded remote entanglement generation}
Heralded remote entanglement generation~\cite{moehring2007entanglement,ritter2012elementary,hofmann2012heralded,bernien2013heralded} can establish entanglement between two local systems that have no direct physical interaction. The effective interaction is realized in the following way: 
\begin{enumerate}
    \item The local systems emit photons which are entangled with themselves. Such matter-photon entanglement can be generally expressed as a standard Bell state.
    \item The two emitted photons are transmitted through a physical channel to an intermediate photonic Bell state measurement (BSM) station.
    \item The photonic BSM realizes an effective entanglement swapping~\cite{pan1998experimental}, which conditioned on success projects the quantum state of two remote systems into a Bell state. 
    \item The classical BSM result is communicated with the two end nodes, which heralds whether the remote entanglement generation is successful.
\end{enumerate}
The above processes are visualized in Fig.~\ref{fig:preliminaries}(a). 

\subsection{Nonlocal operations with entanglement}
In DQC, nonlocal multi-qubit gates between QPUs can be implemented through teleportation. We may either teleport qubits~\cite{bennett1993teleporting} from one QPU to another QPU, perform local multi-qubit gates and teleport the qubit back, or directly teleport the multi-qubit gates~\cite{gottesman1999demonstrating}. These approaches require Bell pairs as resources that are consumed for each attempt, local operations on each QPU, and potential feedforward Pauli frame correction. The circuit for quantum state teleportation is depicted in Fig.~\ref{fig:preliminaries}(b). Fig.~\ref{fig:preliminaries}(c) shows the specific circuit that teleports a CNOT gate. 

\subsection{Related work}
Related to this work, AutoComm~\cite{wu2022autocomm} and QuComm~\cite{wu2023qucomm} are state-of-the-art approaches which take into account quantum circuit structure to harness the benefits of DQC. However, these works ignore one important feature - the probabilistic nature of EPR pair generation. AutoComm identifies burst communication patterns in the input program and utilizes state or CNOT teleportation to allow remote operations. It compiles the circuit and schedules remote operations offline, making it unaware of the real-time EPR pair generation pattern. QuComm identifies multi-node collective communication, and uses ``buffer'' qubits which store generated EPR pairs while communication qubits are freed up to be able to keep generating EPR pairs. 

Several research papers have sought to reduce the communication overhead of distributed quantum programs by exploring various qubit partitioning and mapping techniques~\cite{zomorodi2018optimizing,andres2019automated_partition,davarzani2020dynamic,daei2020optimized_partition,ferrari2021compiler,diadamo2021distributed,dadkhah2022reordering_partition,ferrari2023modular}. Different from the mapping algorithms for monolithic devices~\cite{li2019tackling, niu2020hardware, liu2022NASSC, tan2020optimal, lin2023scalable, liu2023tackling, zou2024lightsabre}, some work~\cite{baker2020time_multicore, bandic2023mapping_multicore, escofet2024revisiting_multicore} focuses on tackling the mapping problem in multi-core quantum computers. Additional work incorporates a recent trend and use reinforcement learning for DQC compiling~\cite{promponas2024compiler,russo2024attention}. These approaches are orthogonal to our work. Finally, a hardware-focused body of work recently emerged as well~\cite{ang2024arquin,kim2024fault,chu2024titan,bahrani2024resource}.


\section{Proposed Architectural Framework}\label{sec:framework}
To motivate our proposed architecture, we first analyze heralded remote entanglement generation for DQC. Then we explain the three principles of our architectural design. 

\subsection{Technical considerations of remote entanglement generation}
The runtime of DQC in realistic heralded remote entanglement generation is affected by the success probability and the cycle time per attempt. The final fidelity of the DQC circuit output is affected by the quality of the generated entanglement (fidelity of the generated Bell pair). However, entanglement infidelity only injects errors into the circuit locally when implementing remote gates, similar to other noisy local gates.

\noindent \textbf{Success probability per attempt:} 
The success probability of a heralded remote entanglement generation attempt is determined mainly by the following factors:
\begin{enumerate}
    \item Local photon-qubit entanglement generation probability $p_{pq}$ during one entanglement generation cycle (attempt).
    \item Photon transmission loss. For typical low-loss optical fiber the efficiency $\eta_t=\exp(-L/L_\mathrm{att})$ is a function of channel length $L$ where the characteristic attenuation length for optical fiber is $L_\mathrm{att}\approx 20~\mathrm{km}$.
    \item In practice photon BSMs are typically implemented using linear optical components whose maximal success probability is 1/2~\cite{lutkenhaus1999bell}. In addition, there are other factors such as optical coupling efficiencies and detector efficiencies. We denote the total success probability of the photon BSM as $p_s$.
\end{enumerate}



Thus, the success probability of one entanglement generation attempt is $p_\mathrm{succ} = p_{pq,1}p_{pq,2}\eta_{t,1}\eta_{t,2}p_s\leq 1/2$. Note that both sides must succeed with the transmission.

\noindent \textbf{Cycle time per attempt:} 
The following three processes are the main contributors to the cycle time:

\begin{enumerate}
\item For realistic DQC, the allowed waiting time for photon emission should be a fixed value (cutoff time) otherwise the probabilistic nature of emission might lead to an excessively long wait for the emission. On the other hand, the chosen cutoff time will affect the photon emission probability for every attempt of the protocol. 

\item If successfully emitted, photons are transmitted via optical channels in which photons travel at the speed of light. The speed of light in optical fiber is $2\times 10^8\mathrm{m/s}$. Assuming that the length of the optical fiber from one QPU to the central BSM station is roughly 10m in a DQC center, the on-way transmission time is then $\sim50$ns. 

\item Latency is introduced when extracting, processing, and transmitting  classical measurement results from detectors. Even after obtaining and processing the measurement outcomes, latency may be introduced when transmitting the classical feedback using classical communication protocols. The lower bound for this last step is the physical transmission of information at the speed of light.
\end{enumerate}

As a result, the total duration (cycle time) $T_\mathrm{EG}$ of a single heralded remote entanglement generation attempt is much longer than the cycle times for local operations $T_\mathrm{local}$. For the typical case we assume $T_\mathrm{EG}\geq 10T_\mathrm{local}$~\cite{wu2023qucomm}.

\subsection{Entanglement buffering}
\begin{figure}[t]
    \centering
    \includegraphics[width=0.9\linewidth]{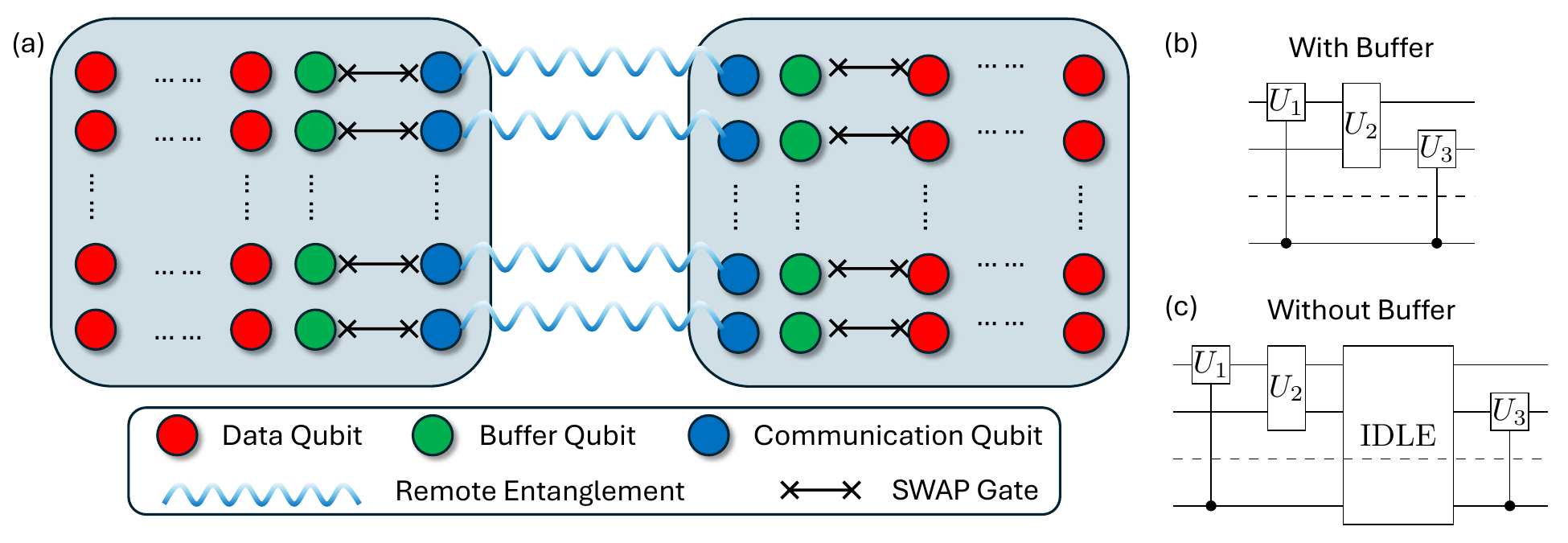}
    \caption{Schematics of the DQC hardware architecture. (a) Example two-node scenario. (b) Example circuit segment which includes remote gates with buffer qubits. (c) Without a buffer qubit there is additional idling before a remote gate can be implemented due to the wait for entanglement generation.}
    \label{fig:buffer}
\end{figure}

We advocate the use of three types of qubits: buffer qubits, data qubits used to evaluate quantum circuits, and communication qubits which generate entanglement. A schematics of the architecture involving the three different types of qubits is shown in Fig.~\ref{fig:buffer}(a). After a pair of communication qubits successfully generates entanglement, each host node applies a local SWAP gate between its communication qubit and a buffer qubit. This results in storing the entangled state in the buffer qubit, whereas the communication qubit is freed up and can continue to be used in entanglement generation attempts. The buffer qubits can store multiple entanglement links to fulfill the demand for remote gates when needed. Decoupling EPR pair generation from storage also enables the pre-initialization of EPR pairs prior to program execution. As demonstrated in Section~\ref{sec:results}, pre-initialized EPR pairs in the buffer (the \texttt{init\_buf} design) reduce overall circuit latency. We note that QuComm~\cite{wu2023qucomm} also formalized the concept of buffering in DQC to ensure sufficient communication resources for collective communication. In contrast, our work emphasizes the critical role of buffering in addressing the challenges of probabilistic EPR pair generation and evaluates the effectiveness of buffer qubits in this context. Next we provide additional motivation for introducing buffer qubits with focus on the system design perspective.

\noindent \textbf{Layering:}
The introduction of buffer qubits in addition to communication qubits and data qubits allows separation of the entanglement generation process in DQC. With buffer qubits, entanglement generation can be implemented as a service that runs continuously~\cite{kolar2022adaptive,zang2024analytical,zhan2025design} in the background. This hierarchical structure is beneficial for modularization of the DQC hardware architecture, thus simplifying the maintenance of each module. On the other hand, the interaction between communication qubits and buffer qubits to implement the SWAP gate can be realized with quantum interconnect technologies~\cite{awschalom2021development,awschalom2022roadmap}.

\noindent \textbf{Multiplexing:}
By attempting entanglement generation in parallel (multiplexed) using multiple available communication qubits, the latency for receiving an EPR pair is reduced due to the increased effective entanglement generation rate. However, it is unlikely that the entanglement links are generated exactly at the time when the remote gates request them, so they must be stored. If there are no buffer qubits, the communication qubits have to also serve as quantum memories. This means that fewer communication qubits are available and the effective entanglement generation rate decreases. Moreover, an insufficient number of available communication qubits can lead to unnecessary idling due to having no available entanglement links, leading to qubit state degradation and longer runtimes. Such a scenario is illustrated in Fig.~\ref{fig:buffer}(b) and (c).

\subsection{Asynchronous remote entanglement generation}
\begin{figure}[t]
    \centering
    \includegraphics[width=0.9\linewidth]{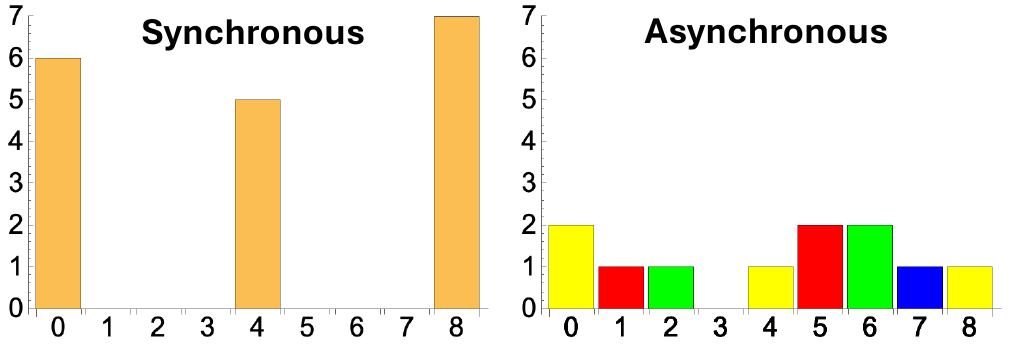}
    \caption{Example entanglement generation patterns in the time domain. Synchronous and asynchronous attempts between each communication qubit pair are depicted. The vertical axis denotes the number of entangled pairs generated in one time unit, and the horizontal axis denotes time.}
    \label{fig:asynchronous}
\end{figure}

The cycle time for remote entanglement generation is generally much longer than the time scale of local quantum operations. For this reason we propose the use of asynchronous entanglement generation among all available communication qubits, a process that is able to ``smoothen'' the entanglement generation temporal pattern. This yields better resource allocation in the time domain.

In Fig.~\ref{fig:asynchronous}, we visualize the number of successfully generated entanglement links over time, in units of local operation cycle time $T_\mathrm{local}$, and without loss of generality we have assumed $T_\mathrm{EG}=4T_\mathrm{local}$. In the left panel, when all entanglement generation attempts are synchronous, the entanglement links will arrive in bursts. Remote gates thus need to wait for the burst arrival of entanglement links. After some links in the burst are consumed, the remaining links have to be stored in buffer qubits, which leads to idling decoherence, thus lowering the fidelity of remote gates. In contrast, the right panel demonstrates the pattern of asynchronous attempt when communication qubits are divided into 4 sub-groups, whose starting times of entanglement generation attempts are separated by $T_\mathrm{local}$. Different colors denote different sub-groups. In this way, the number of entanglement generated within one time unit is lower, but they distribute more uniformly over time so that some remote gates are able to utilize the entanglement links as soon as they are generated. Additionally, the burst arrival of entangled states can lead to excessive waste when we apply a cut-off policy to buffer qubits, i.e. reset buffer qubits if they store entanglement for too long to avoid too much decoherence in entangled states.

\subsection{Adaptive scheduling of remote gates}
\begin{figure}[t]
    \centering
    \includegraphics[width=0.9\linewidth]{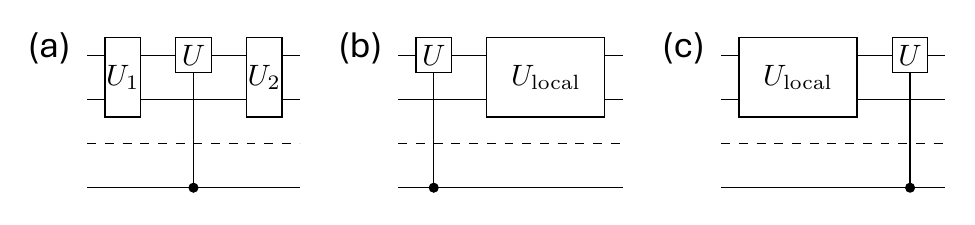}
    \caption{Example circuit segment variants for different scheduling strategies. (a) Original circuit that does not consider the entanglement generation pattern. (b) ASAP implementation of the remote gate. (c) ALAP implementation of the remote gate.}
    \label{fig:schedule}
\end{figure}

Adaptive scheduling of remote gates represents an additional optimization opportunity that can be implemented in software. Note that for simplicity this work assumes all remote operations are implemented through gate teleportation, and we leave the simultaneous inclusion of both state and gate teleportations as a future work.

Information about existing generated entanglement links can be exploited by the DQC controller. Before a remote gate is executed, we examine whether there are already enough entanglement links stored in the buffer qubits. If yes, we can execute the remote gate earlier so that it can consume existing entanglement as-soon-as-possible (ASAP). This will create a longer time interval between the executed remote gate and subsequent remote gates, and the communication qubits will have a higher chance to create additional entanglement links demanded by the subsequent gates. If there are not enough entanglement links available yet, we can push the remote gate back (as-late-as-possible, ALAP) so that there is more time to generate the required entanglement.

This adaptive circuit modification process must be performed in real time. Because dynamically resynthesizing the quantum circuit based on the number of available EPR pairs $e$ in the buffer is difficult, we can instead statically pre-compile multiple versions of the circuit. To ensure the scalability of the classical compilation, instead of recompiling the whole circuit, we partition the circuit into segments and use a look-up table strategy to manage them. Each segment is compiled with different scheduling policies, and the appropriate circuit version is selected in real time based on the number of available EPR pairs $e$. To achieve this, we first identify the remote gates and partition the original circuit into segments, where each segment contains $m$ remote gates. The parameter $m$ is tunable and is set to the product of the number of communication qubits and the EPR pair generation rate $p_\mathrm{succ}$ in our experiments.
After partitioning the circuit, we can obtain the variants of the circuit segment corresponding to ASAP and ALAP, respectively, as illustrated by the examples in Fig.~\ref{fig:schedule}. During execution, if $e > m$, the controller looks up the next segment with ASAP policy. If $e = 0$, the controller opts for the ALAP policy. Otherwise, the controller uses the original scheduling. We leave more complicated scheduling strategies for future work.

\section{Evaluation Methodology}\label{sec:methodology}
In this section we describe our performance evaluation benchmarks, system configurations, comparison baselines, and figures of merit. We also describe the entanglement generation dynamics simulation.

\subsection{Benchmarks and Configuration}\label{sec:benchmark}
\noindent\textbf{Benchmarks:} We evaluate 6 benchmarks that span different problem sizes and applications. The benchmarks are selected based on the proportion of remote gates in their circuits. The 1D Transverse-Longitudinal Ising Model (TLIM) circuit~\cite{sopena2021simulating} features linear connectivity and a small number of remote gates. The Quantum Approximate Optimization Algorithm (QAOA)~\cite{farhi2014qaoa} is used to solve the MaxCut problem on regular graphs of degrees 4 and 8, and these circuits therefore have  a medium proportion of remote gates. For instance, QAOA-r4-32 represents a 32-qubit QAOA circuit designed to solve the MaxCut problem on a degree-4 regular graph. Finally, the Quantum Fourier Transform (QFT)~\cite{coppersmith2002qft} benchmark requires full connectivity and exhibits a high proportion of remote gates. The benchmark properties are listed in Table~\ref{table:benchmarks}. The code and data that support the findings of this work are available upon request from the authors.

\begin{table}[htbp]
\centering
\caption{Benchmark properties}
\label{table:benchmarks}
\resizebox{\linewidth}{!}{
\begin{threeparttable}
\begin{tabular}{|c|c|c|c|c|c|}
\hline
Name & \#qubits & \#local 2Q\tnote{a} & \#remote 2Q\tnote{a} & \#1Q\tnote{a}  & depth\\ \hline
TLIM-32 & 32 & 300 & 10 & 640 & 40\\ \hline
QAOA-r4-32 & 32 & 52 & 12 & 64 & 21\\ \hline
QAOA-r8-32 & 32 & 91 & 34 & 64 & 64\\\hline
QFT-32 & 32 & 240 & 256 & 32 & 63\\ \hline
QAOA-r4-64 & 64 & 104 & 28 & 128 & 24\\ \hline
QAOA-r8-64 & 64 & 174 & 82 & 128 & 84\\\hline
\end{tabular}
\begin{tablenotes}
    \small
    \item[a]\#local 2Q denotes the number of local two-qubit operations,
    \#remote 2Q denotes the number of remote two-qubit operations, and
    \#1Q denotes the number of single-qubit operations
\end{tablenotes}
\end{threeparttable}
}
\end{table}

\noindent\textbf{System Configuration:} The properties of the different types of quantum operations are listed in Table~\ref{table:backend_specification}. The operation latencies listed in the table agree with those in~\cite{wu2023qucomm}. In our evaluations, we assume that the success probability of one round of entanglement generation is $p_\mathrm{succ}=0.4$, the backend has a decoherence time $1/\kappa=150\mu\text{s}$ where $\kappa$ is the decoherence rate, and the local CNOT gate time is $300ns$.

\begin{table}[htbp]
\centering
\caption{Quantum Operation Properties}
\label{table:backend_specification}
\begin{threeparttable}
\begin{tabular}{|c|c|c|c|c|c|}
\hline
Name & Latency & Fidelity\\ \hline
1Q gates & 0.1 & 99.99\%\\ \hline
Local CNOT gates & 1 & 99.9\%\\ \hline
Measurement & 5 & 99.8\%\\ \hline
EPR pair preparation & 10 & 99\%\\\hline
\end{tabular}
\end{threeparttable}
\end{table}

\noindent\textbf{Baseline:} Similar to prior work~\cite{davis2023towards}, we utilize the METIS partitioning solver~\cite{karypis1997metis} to determine the partitions that are be assigned to different nodes, aiming to minimize the number of remote operations.

\subsection{Figures of merit}
We evaluate different designs by comparing the circuit depth and the circuit fidelity. The circuit depth represents the total latency of the circuit. A depth of one corresponds to the latency of a single local CNOT gate. The circuit fidelity represents the quality of the final output. The fidelity is estimated as the product of fidelities of all local single-qubit gates, local two-qubit gates, remote gates implemented through gate teleportation, and an additional idling decoherence factor which accounts for latency. The local gate fidelities we use are listed in Table~\ref{table:backend_specification}. It is important to note that the fidelity of remote gates depends on the fidelity of the consumed entanglement link, which is affected by the buffer qubit decoherence. We use the phenomenological exponential decay factor $\exp(-\kappa t)$, where $t$ is the idling time, to model idling decoherence.

\subsection{Simulation of remote entanglement generation}
Inspired by a recent open-source ad hoc quantum network simulation~\cite{zang2023entanglement}, we simulate asynchronous entanglement generation by communication qubits and storage of entanglement by buffer qubits. The system parameters are documented in Sec.~\ref{sec:benchmark}. We assume that the initially generated Bell state is in the Werner form, i.e., a mixture of a pure Bell state and a 2-qubit maximally mixed state. We also assume that all buffer qubits have an identical decoherence rate, and the decoherence channel is the unbiased depolarizing channel, i.e. the three Pauli errors have an identical probability. As a result, the idling dynamics of the Bell state fidelity is given by $F(t) = F_0\exp(-2\kappa t) + [1 - \exp(-2\kappa t)]/4$, where $F_0$ is the initial fidelity. The fidelity of a remote gate is obtained through the evaluation of the gate teleportation circuit which includes a noisy Bell state, noisy local 2-qubit gates, and a noisy single-qubit measurement. 

\section{Results}\label{sec:results}
In this section, we evaluate the DQC circuit depth and fidelity across various architecture designs and system sizes. The designs we consider include \texttt{original}: without any buffer qubits, \texttt{sync\_buf}: with buffer and synchronous EPR pair generation, \texttt{async\_buf}: with buffer and asynchronous EPR pair generation, \texttt{adapt\_buf}: with asynchronous EPR pair generation and adaptive remote gate scheduling, \texttt{init\_buf}: with pre-initiated EPR pairs in buffers, and \texttt{ideal}: execution on a monolithic device without remote operations. All reported results represent the average of 50 runs.

\subsection{Comparison of different designs}
\begin{figure}[htbp]
\centerline{\includegraphics[width=0.95\linewidth]{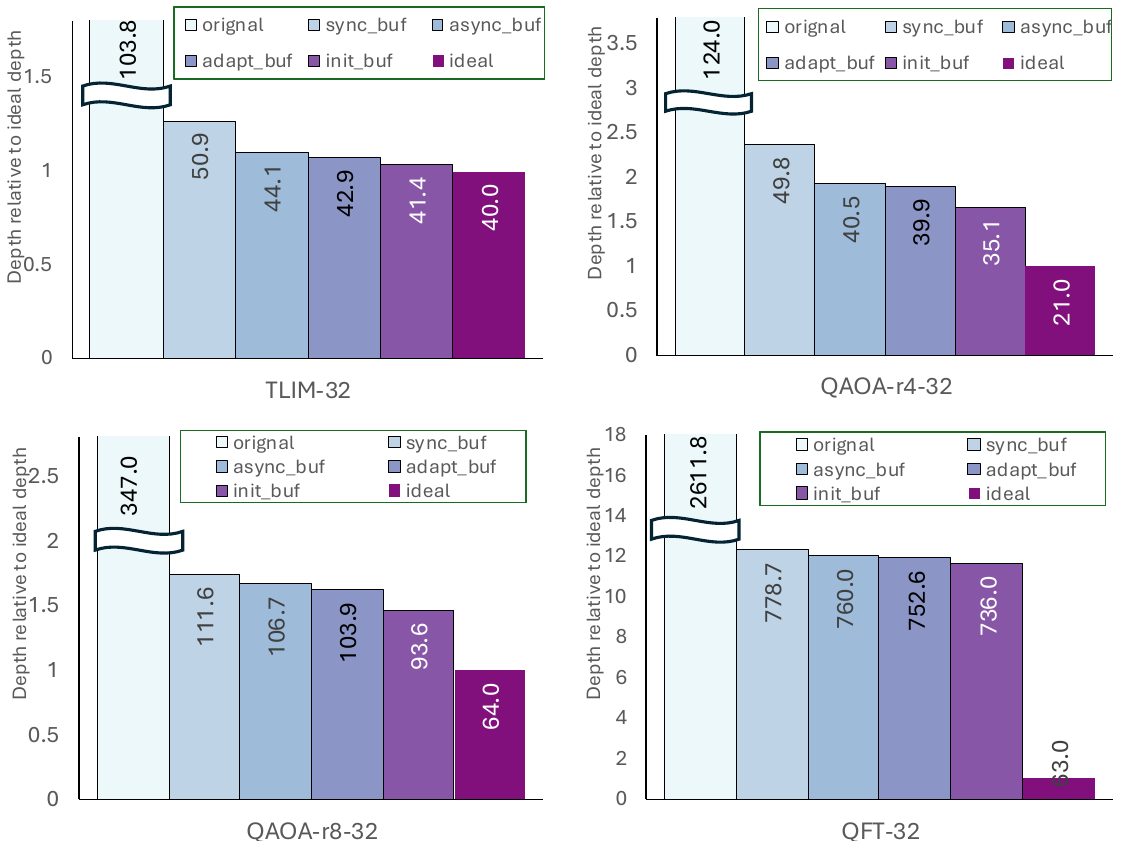}}
\caption{Comparison of circuit depths across benchmarks and designs.}
\label{fig:32q_depth} 
\end{figure}

\begin{figure}[htbp]
\centerline{\includegraphics[width=0.95\linewidth]{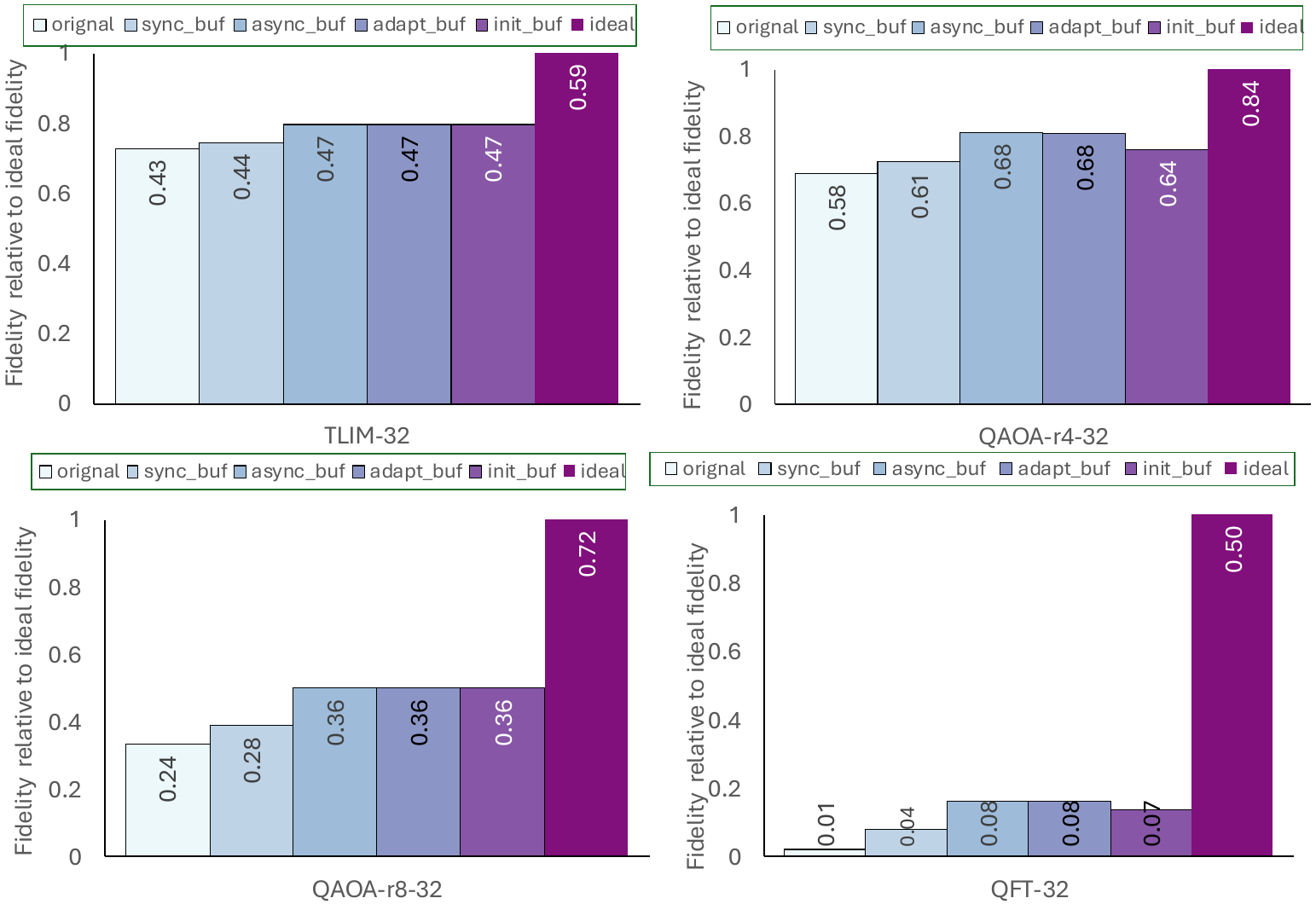}}
\caption{Comparison of circuit fidelities across benchmarks and designs.}
\label{fig:32q_fidelity} 
\end{figure}

We evaluated different designs on a simulated 2-node 32-data-qubit DQC architecture with 16 fully-connected data qubits assigned to each node. Each node contains 10 additional communication qubits and 10 buffer qubits. The results are presented in Figure~\ref{fig:32q_depth} and~\ref{fig:32q_fidelity}. In Figure~\ref{fig:32q_depth}, we compare the circuit depth across different designs. The y-axis represents the circuit depth relative to the ideal circuit depth. The \texttt{original} design, which lacks buffer qubits, results in significant EPR pair waste and a substantial increase of the circuit depth. As shown in the figure, the largest reduction of the depth is achieved by leveraging buffer qubits. The \texttt{sync\_buf} design reduces the circuit depth by $61.7\%$. However, the synchronous EPR pair generation pattern in \texttt{sync\_buf} does not align well with the remote gate request pattern. The asynchronous EPR pair generation design \texttt{async\_buf} yields an additional average $7\%$ reduction of the circuit depth. Building on this, the adaptive scheduling \texttt{adapt\_buf} further reduces the depth. By combining the pre-initialized EPR pairs and the adaptive scheduling, \texttt{init\_buf} achieves an additional $7.5\%$ depth reduction compared to the non-adaptive \texttt{async\_buf} design.

We also compared the estimated circuit fidelity across different designs. As shown in Figure~\ref{fig:32q_fidelity}, the \texttt{async\_buf} and \texttt{adapt\_buf} designs achieve the same fidelity, yielding an average improvement of 2$\times$ and 1.32$\times$ compared to the \texttt{original} and \texttt{sync\_buf} designs, respectively. Although the \texttt{init\_buf} design leverages the pre-initiated EPR pairs to shorten the circuit depth, the extended idle time of these EPR pairs can lead to reduced fidelity.

\subsection{Impact of the number of communication qubits}

\begin{figure}[htbp]
\centerline{\includegraphics[width=0.95\linewidth]{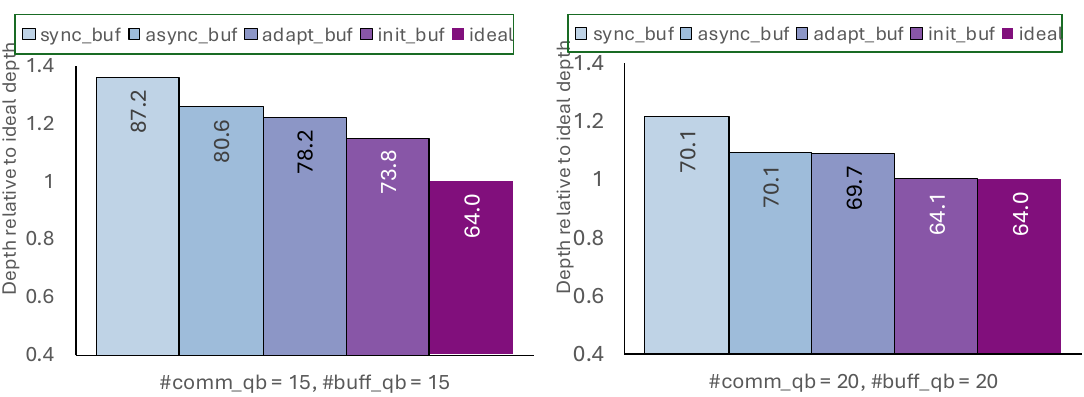}}
\caption{Circuit depth of QAOA-r8-32 with different numbers of communication and buffer qubits.}
\label{fig:comm_qb} 
\end{figure}

We evaluate performance with a different number of communication and buffer qubits. The results are shown in Figure~\ref{fig:comm_qb}. We pick the QAOA-r8-32 benchmark since the TLIM benchmark already achieves a near-ideal depth with 10 communication qubits, whereas the QFT-32 benchmark involves an excessive number of remote gates. The \texttt{original} design is excluded as it shows minimal change, allowing us to focus on the optimized designs. This also emphasizes that merely increasing the number of communication qubits is ineffective, as EPR pairs are wasted without proper storage in the buffer. As we increase the number of communication and buffer qubits, we notice a significant reduction of the circuit depth. Among the designs, \texttt{init\_buf} consistently delivers the best performance. When the number of communication qubits reaches 20, it achieves a near-ideal depth, indicating that all remote gates are served immediately. Despite the significant reduction of the circuit depth, the circuit fidelity remains almost unchanged. Hence, we opted to not include the figure. The primary reason is that with asynchronous and adaptive scheduling EPR pairs are consumed immediately after generation, maintaining high fidelity across different cases. The relatively short circuit depth also minimizes the impact of decoherence errors, resulting in negligible differences of the fidelity.

\subsection{Evaluations of larger systems}

\begin{figure}[htbp]
\centerline{\includegraphics[width=0.95\linewidth]{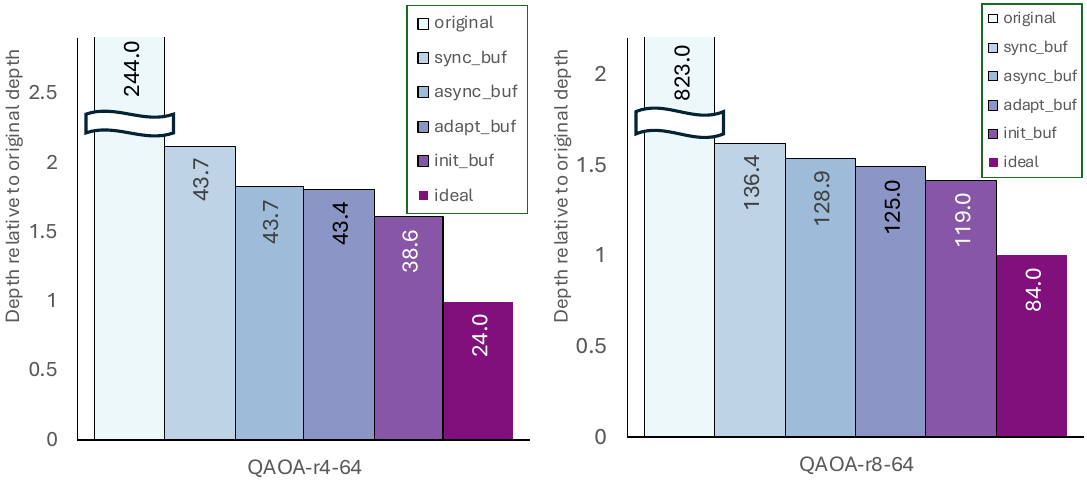}}
\caption{Comparison of circuit depths across benchmarks and designs on a 64-qubit system.}
\label{fig:64q_depth} 
\end{figure}

We evaluate two QAOA benchmarks on a 2-node 64-data-qubit system with 32 data qubits allocated to each node. Each node contains 20 additional communication qubits and buffer qubits. The circuit depth comparison is presented in Figure~\ref{fig:64q_depth}. The \texttt{init\_buf} design achieves a $12\%$ circuit depth reduction compared to the \texttt{sync\_buf} design. As illustrated, our proposed designs continue to significantly reduce the circuit depth for larger system sizes.

\section{Conclusion and Discussion}\label{sec:conclusion}
In this paper, we study a new DQC architecture that leverages hardware and software co-design. One of our key observations is that buffering of successfully generated entanglement significantly reduces the circuit latency. Additionally, we introduce an asynchronous entanglement generation strategy that optimizes resource allocation and an adaptive scheduling mechanism that dynamically responds to real-time EPR pair availability. These advancements collectively enhance the performance and scalability of DQC systems, paving the way for more practical and efficient implementations that support the needs of real-world applications.

\section*{Acknowledgments}
This material is based upon work supported by the U.S. Department of Energy, Office of Science, National Quantum Information Science Research Centers. This material is also based upon work supported by the DOE-SC Office of Advanced Scientific Computing Research MACH-Q project under contract number DE-AC02-06CH11357. A.Z. and T.Z. acknowledge support from the NSF Quantum Leap Challenge Institute for Hybrid Quantum Architectures and Networks (NSF Award 2016136).

\bibliographystyle{IEEEtran}
\bibliography{references}

\end{document}